# Blurring Boundaries:

## Toward the Collective Empathic Understanding of Product Requirements

Robert C. Fuller
Electrical and Computer Engineering
The University of British Columbia
Vancouver, Canada
e-mail: rfuller@ece.ubc.ca

Philippe Kruchten
Electrical and Computer Engineering
The University of British Columbia
Vancouver, Canada
e-mail: pbk@ece.ubc.ca

*Abstract*— **Within the agile paradigm, many software product companies create cross-functional product development teams that own their product or a defined set of product features. In contrast to development teams operating within a heavily-disciplined software development process, these product teams often require a deeper and, importantly, a collective understanding of the product domain to serve as a rich context within which to understand the product requirements. Little is known about the factors that support or impede these teams in collectively achieving this deep understanding of the product domain. Using Constructivist Grounded Theory method, we study individuals and teams across seven software companies that create products for a diverse range of markets. The study found that certain organisational and planning process factors play a significant role in whether product development teams have the potential to collectively develop deep domain understanding. These factors also impact individual and development team dynamics. We identify two essential metaphorical dynamics of broadening the lens and blurring boundaries that cross-functional product teams employ in order to fully embrace product ownership, visioning, and planning towards achieving a deep collective domain understanding, creating a richer context in which to understand product requirements. We also conclude that the highly specialised nature of many organisational models and development processes is contraindicated for cross-functional product development teams in achieving this deep collective understanding and we call for a rethinking of conventional organisational and product planning practices for software product development.**

### 1. INTRODUCTION

How well do your software product development teams grok the world their products are intended for? In Robert A. Heinlein's famous 1961 science-fiction novel, Stranger in a Strange Land (Heinlein, 1961), Mahmoud explained that to "grok means to understand so thoroughly that the observer becomes a part of the process being observed" and he declared, "I am all that I grok". Later, not to ignore a pop culture word that was not going away, the Oxford English Dictionary defined the verb grok as to "understand (something) intuitively or by empathy" (Oxford English Dictionary, 1989). This article is about grokking, it is about collective grokking, and it is about collective grokking by software product development teams of the domain for which their products are intended for the purpose of creating a rich context within which to understand product requirements and develop the software - 'We are all that we grok'.

Successful software products are almost always a result of creativity, innovation, and vision that are all created and maintained throughout the entire product life cycle. As noted by Hoegl and Gemuenden (2001), teamwork quality and the success of innovative projects is a measure of cohesion, mutual support, and coordination, thus successful software products are more likely to be created by strong, healthy, cohesive teams than they are by workgroups, simple assemblies of talented individuals. It is clear by observation from our long industry experience that strong software development teams vary considerably in their ability to collectively grok their product's domain. If the reasons for this variability were better understood, software development practitioners could use that knowledge to proactively influence the factors at play - factors that impact the success of their software product teams, the success of the software products, and the success of their companies.

This study, motivated by our direct industrial experience, examines factors that impact software development teams' ability to achieve a deep, collective understanding of the product domain (and, hence, the context of the product requirements).

We found that the organisational and product planning process models that software product development teams operate within have a strong influence on the dynamics of the software product development teams and their success at *Blurring Boundaries* towards collectively grokking the product domain. In this article, we discuss the impacts in more detail, impacts which often go unnoticed by leaders. At other times, especially when the impacts are negative, they become the proverbial 'elephant in the room', known but deliberately overlooked due to an inability to correct the conditions.

This study also highlights behaviours and dynamics that tend to occur within cross-functional product teams in response to these organisational and product planning process models.

Early results of this research (Fuller, 2019a) highlighted that the organisational model surrounding the cross-functional development teams (that is, the composition of functional departments and the manner in which teams are formed) had a significant influence on the teams themselves – on the intra-team communication dynamics, on both the individual and collective sense of ownership of and commitment to the product vision/goals/plans, and on the collective capability of the team to deeply understand the product domain.

Furthering those results, we use Charmaz's Constructivist Grounded Theory method (2014) (described further in Chapter 6) to examine a more comprehensive view of the dynamics and context of cross-functional product teams (CFPTs), including the teams' role in software product visioning and complex planning and in the teams' interest, capability, and efforts to grok the product domain.

Our main contribution is our finding that teams that strive to collectively grok the product domain tend to be ones that have more team cohesion with a clear and broad ownership of their product. This also contributes to the shortfall of considerations in the Requirements Engineering discipline of the role of human factors. We also call for a reconsideration of the common mechanistic view of organisational structures and process models in software product development.

The remainder of this paper is structured as follows: Chapter 2 – Background provides an overview of the historical background of the topic. Chapter 3 – Research Question and Motivation frames the reason for the research and clarifies the question. Chapter 4 – Research Focus and Challenges describes what we are aiming to achieve and a brief description of the research scope. Chapter 5 – Related Work positions this study with respect to three related areas of research while Chapter 6 – Method of the Research overviews the research methodology and activities conducted throughout the study. Chapter 7 – Observations and Analysis describes the findings which are discussed further in Chapter 8 – Discussion, where we provide deeper analysis of the observations. Chapter 9 – Implications for Practice discusses the research findings in terms relevant for industrial practice, and finally Chapter 10 - Conclusion and Future Work, offering thoughts about the research contribution thus far and possible directions for future research.

## 2. BACKGROUND

As the 20th century drew to a close, three forces had taken hold which dramatically changed the nature and challenge of software product development. Bill Gates has been quoted as saying that Microsoft was founded with a vision of "a computer on every desk, and in every home". The first force that we refer to is that vision becoming a reality with the computer industry having broken through a price/performance threshold for personal computers which brought computing capability to almost every desk in the workplace and to almost every home. This created a significantly larger and more diverse demand for software.

The second force was the widespread introduction of graphical user-interfaces (GUIs), primarily fueled by the Macintosh and Windows operating systems, which dramatically enriched the possibilities (and complexities) for human-computer interaction (HCI), opening up new dimensions for computer user in both the workplace and in the home.

Finally, the third force was the rise of the Internet, which allowed for the emergence of totally new business models by introducing entirely new possibilities for accessing data and for using information technology as we knew it then (Wasserman, 2011).

In addition to these forces increasing the 'art of the possible' in software (as well as significantly increasing the complexity of software design, development, and testing), these forces also resulted in more software being developed as products for a market (Wasserman, 2011) with potential customers instead of software being developed predominantly as bespoke solutions for known customers (including internal application development). This latter model had been the mainstream context for software development prior to these forces taking hold. This shift towards more product software development was significant because it alone introduced even more risk and uncertainty into the entire software development process.

This had all the signs of a crisis period and, in response to this, a Kuhnian "model revolution" (Kuhn, 2012) emerged during this crisis period in the software industry that took a new view on change, risk, and uncertainty in software development in general and which was critically important for software product development. This new 'agile' paradigm (Agile Alliance, 2001) accepted that requirements (or understanding of market needs) could (and, probably, would) change throughout the (now product) development life cycle. Not only would the needs likely change but that better and deeper understandings of the requirements would emerge throughout the development effort in contrast to more disciplined Software Development Life Cycles (SDLC) that strived to lock down requirements in the specification and planning stages in order to minimise the uncertainty of timeframes, costs, and deliverables. Taking forms of iterative and incremental approaches to solution development and using cross-functional teams to attempt to 'discover' the needs throughout the development effort, these approaches viewed emergence as a fact of life rather than a failure of the requirements elicitation and analysis activities. "Embrace change" wrote Beck (2004) in his seminal book on eXtreme Programing (XP).

This acceptance of uncertainty and emphasis on learning throughout the development process, placed a new and greater focus on the software development team, recognizing that prescriptive processes were insufficient to ensure project success in these complex and emergent conditions and that the dynamics of the development team, which was now usually cross-functional and often empowered to truly own the software product, and the team's understanding of the problem domain was considered a critical success factor in delivering software. The statement from the Agile Manifesto, Individuals and interactions over processes and tools (Agile Alliance, 2001) rang loud for those adopting this new paradigm for software development.

Out of this paradigm shift emerged an entirely new challenge for software development leadership, one that many were ill-prepared for, namely, how to lead teams and teams-of-teams successfully under these conditions of rapid change and uncertainty without a prescriptive development methodology to guide them.

While the agile approaches that emerged improved many of the issues that were breaking down during the crisis period, they fell silent on the issue of the development team's understanding of the problem domain, choosing to still refer figuratively to a 'customer', whether one actually existed or not. In this regard, even today, many software product organisations still operate as if they are developing for a single customer. And, when they are not, they often anoint an internal surrogate (the so-called "customer on-site" of XP), an authoritative voice that the development team can iteratively interact with to clarify requirements and validate results. These internal roles may hold various titles such as Product Owner, Product Manager, Market Analyst, Customer Researcher, etc.

However, as software solutions addresses more complex and subtle needs, as software technology continues to become increasingly complex, and as software development becomes often more product development intended for a whole market rather than a single customer, a critical challenge emerges: namely how software development teams gain a deep understanding of the world for which their product is intended, an understanding that cannot be passed on to the team by any single voice, much less an internal one. Certainly, there are analysis techniques, e.g. from marketing and design thinking, to 'hear' the market and learn about it. These are helpful, however market participants have tacit knowledge, as Polyani (2009) states "people know more than they can tell," and they know more than can easily be observed. A form of this problem commonly occurs with the popular 'user story' technique of communicating end-user requirements (Cohn, 2004) when it's later discovered that the story doesn't reflect an actual need but rather simply an articulation of what someone wants, resulting in, "I know that's what I said I wanted but that doesn't seem to be what I need." -- they know more than they can tell. Sometimes, they can't even express what they want at all, resulting in IKIWISI (I'll Know It When I See It). Boehm (2000) described an additional form of IKIWISI where the customer initially thinks they know what their needs are but their understanding of those needs changes over time with continued use.

In early times, requirements were less complex. The available technology significantly constrained what was possible, needs could be more precisely, unambiguously, and completely expressed, and quite often the requirements came from an identifiable customer, techniques such as having at least one domain expert on (or available to) the team were often sufficient. Today, however, with much more technical and problem complexity, heterogenous customer targets, opaque markets, competitive uncertainties, etc., it is insufficient to simply have access to one person with this deep understanding. And it is even less sufficient to have this deep understanding residing outside the development team. Yet, many software development organisations continue to isolate their development teams in this respect, communicating requirements to the development team as desiderata, and often resulting in requirements fixation (Mohanani et al., 2014) and achieving disappointing results.

Rather, it is important that everyone on the development team has a deep domain understanding and it is critical that everyone understands it in a compatible and consistent way. This is because team members (individually, in sub-teams, and across all functional roles) make decisions continually throughout product design and development. Indeed, they make decisions continually throughout the entire product life cycle, based on their individual understanding of the context of the requirements, and much of that context understanding is tacit. This challenge is well expressed by Berry (1995) when discussing assumptions in requirements engineering amongst team experts:

"It seems that among experts, a common disease is the presence of unstated assumptions. Because they are unstated, no one seems to notice them. Worse ... it seems that no two people have the same set of assumptions, often differing by subtle nuances that are even more tacit than the tacit assumptions. It is these assumptions that confound attempts to arrive at consensus, particularly because none of the players is even consciously aware of his or her own assumptions and certainly not of the differences between the players' assumptions" (p.180)

Thus, it behooves product development teams to strive for a deep collective understanding of the context of their product, a shared mental model of all the elements of the domain, since many decisions made throughout the development life cycle will be made unconsciously within the team's understanding of the domain context.

In other words, it behooves them to grok to the best of their ability. When we use the term grok in this article, we are referring to cognitive empathy, coupled with skilled perspective-taking. Specifically, we use the definition of cognitive empathy to be "the ability to imaginatively step into another domain, understand the perspectives of those in that domain, and use that understanding to guide decisions" (Krznaric, 2014). Increasingly, the success of software product development teams, indeed of the companies themselves, depends on the degree to which the team collectively groks not only the product requirements themselves but also, and very importantly, the context for those product requirements, and then is able to use that collective grokking of the requirements and context to guide their development decisions.

## 3. RESEARCH QUESTION AND MOTIVATION

From our extensive leadership experience in the software industry, we observed that cross-functional product teams (CFPTs) achieve significantly different degrees of success in collectively grokking the product domain, even when they are sharing the same organisational, process, and leadership environments. Software development practitioners have no theories that help explain why. Without explanatory models, industry leaders are unable to proactively nurture conditions and relevant factors to support teams to be as successful in this regard as they could be.

The overall purpose of this qualitative research study is to develop a substantive theory that answers the following general question:

> "How do cross-functional software product teams collectively achieve a deep and shared understanding of the product domain?"

Insights gained from this study will help industry practitioners explain why certain prevailing techniques and empirical approaches for understanding software solution needs are often inadequate, why some succeed while others do not. It is also intended to offer interpretive insights and guidance into factors affecting how creativity and innovation occurs (or is interfered with) within software product teams.

Early in this study, it became evident that there were differences across participant organisations that were even more pronounced than the differences within any given participant company, suggesting that there might be broader contextual factors at play. Specifically, we observed (Fuller, 2019a) that the organisational model surrounding cross-functional product teams (CFPTs) has a significant impact on the team itself, facilitating or impeding the team's capability to grok. This caused us to expand the scope of the study to examine additional factors which were both internal and external to the teams. As a result, we formulated the following specific research question:

> RQ: "what factors support or impede cross- functional software product teams in collectively achieving a deep and shared understanding of the product domain?"

This article focuses on this specific question, aiming to offer insights into factors that support or impede CFPTs in collectively grokking, achieving a deep understanding of the context of their products.

## 4. RESEARCH FOCUS AND CHALLENGES

As the saying goes, "a fish doesn't know it's in water", and so the intended users of software products often cannot envisage an ideal (or, sometimes, even a conceptually different) solution to their needs. Nor can they often clearly communicate the context in which they operate because they are trapped in that context. So, for software development teams to understand and define that which they cannot easily see, to understand the 'why' more than the 'what', to understand the functionality needs, the supra-functionality needs (attributes that satisfy needs beyond the utilitarian functional needs, including the emotional and cultural relationship between the products and the user), and the context of all those needs, it is necessary somehow for the team to figuratively become one of the those targeted to use the software solution, and to truly learn from that immersion. As we noted earlier, it is insufficient to simply have access to someone with this deep understanding or to have someone on the team that does, the team itself needs to grok, to collectively understand so thoroughly that they figuratively become part of the process they're observing ('We are all that we grok').

This is difficult. This is difficult because it involves somehow blurring the perceived boundaries between the team's world and the target environment (e.g., a small software team and a Fortune 500 marketing department). It is difficult to be an outsider and obtain an insider's perspective and knowledge. It is also logistically messy. It does not easily fit into established software engineering practices nor is it well-supported by traditional software engineers' training. Considering that software solutions are a result of a collaborative cross-functional team effort, this difficulty and messiness is even more acute.

The focus of this research then is practicing software product teams in action. Our interest is in teams with some degree of ownership for the product or for a well-defined subset of one (i.e., not simply satisfying provided specifications). In our study, we also examined teams that are, intentionally or unintentionally, not so empowered. For further contrast, the study included one large software firm that develops bespoke solutions. The study considers empirical adaptations these teams made to established software engineering practices and methods towards furthering their grokking of the context in which their users operate, what we refer to as the 'supra-domain', the business needs, technological, cultural, and political context of the product domain. The research also examines how software development individuals and teams, who are trained and encouraged to apply their best judgement, suspend those judgements and opinions at critical times in order to connect with and exercise empathy for the domain for which their solution is intended.

## 5. RELATED WORK

Prior to commencing the research, we surveyed published material in 3 areas: requirements engineering, design science, and collective sensemaking, disciplines that we felt held some relevance to the research questions.

This research is primarily related to Requirements Engineering (RE), specifically requirement elicitation and validation. Reviewing all the papers at the IEEE International Requirements Engineering Conference over the past decade, as well as many other published papers relating to RE, we found growing sentiments expressed about the challenges and shortcomings of prevailing approaches to RE (e.g., Schon et al., 2017). These prevailing approaches tend to focus on techniques and methods to specify and validate detailed requirements more than on deepening the individual practitioners' and their teams' understanding of the context of the requirements.

There is strong dissatisfaction in the RE field. Some agile development thought-leaders such as Cohn are blunt about it: "The idea of eliciting and capturing requirements is wrong." (2004). While many researchers hold to prevailing views, believing that we just need better techniques to improve effectiveness, others are echoing the tone of Cohn's critique, suggesting that the notion of requirements itself may be counterproductive (e.g. Mohanani et al., 2011; Ralph and Mohanani, 2015; Guinan et al., 1998) or even illusory (Ralph, 2013). This is reflective that some software product development practices still operate in the process-driven paradigm and are experiencing what Kuhn (2012) described as the incommensurability across paradigms - that is, some methods from a process-driven paradigm are not necessarily appropriate outside of that paradigm due to differences in conceptual frameworks. While there are certain domains where the 'techniques and methods' approach is entirely adequate and appropriate, the concerns expressed in the literature cited and our focus in this study is on the majority of product and problem domains that do not lend themselves to complete and unambiguous specifications and, therefore, where a key success factor is the ability of the cross-functional product teams (CFPTs) to achieve a deep understanding of the product domain beyond just what is articulated in the requirements specifications.

This controversy in the RE field and the inherent difficulty in investigating the RE space led to the formation in 2012 of the Naming the Pain in Requirements Engineering (NaPiRE) initiative (NaPiRE, 2020), a large-scale community endeavour run by researchers world-wide which periodically surveys current practices and problems of RE within industry. This initiative is partly motivated by the view that RE research is not sufficiently driven by problems emerging from industry nor is it even sufficiently informed about the state of industrial practice in RE. After several open global surveys, NaPiRE cites the most frequently stated reason for project failure is incomplete requirements. This is despite the fairly widespread utilisation of clear RE process models or artefact templates. NaPiRE goes on to note that "since requirements engineering…is highly human-based, we face the challenge to create a solid empirical basis that allows for generalisations taking into account the human factors that influence the…discipline" (Wagner et al., 2019, p.3) leading to the observation that "there is still a lack of theories in requirements engineering" (Wagner et al., 2019, p.5). Our research is positioned in this condition of a lack of RE theories, incomplete requirements being the main cause of project failure, RE being highly driven by human factors, specifically an increasing dependence on the development team understanding of requirements and their context instead of relying on individual understanding that is passed on or made available to the development team.

The design science and design thinking space has considerable material regarding empathy-driven design (translating human needs to experiences), e.g., (Koppen and Meinel, 2012; van Rijn et al., 2011; Postma et al., 2012; Woodcock et al., 2018; Dong et al., 2018; Kourprie and Visser, 2009; Kolko, 2014). However, we found these works fall short of addressing our research question in three critical respects: 1) the focus on the design activity as an early part of an essentially sequential product development process rather than design as part of an on-going continuous product development effort, 2) the tendency to focus on the design individual or only the design team rather than the entire development team and, 3) even when the design team is considered, it tends not to be viewed as a unit regarding its empathic capability. Hence, the notion of a collective empathic understanding appears to be absent.

Design science models described by Wieringa (2014) acknowledge the challenge that empathy-driven requirements understanding attempts to address. He notes, "stakeholders rarely if ever are able to specify requirements" (2014, p. 52) and "requirements are not answers to questions … they are the results of design choices we make, jointly with, or on behalf of, the stakeholders" (p. 52). Wieringa describes a 'contribution argument' approach to justify the design choices but it falls short of addressing this need of deep understanding in the product context due to the 'project' (as opposed to continual) model. Also, while the approach aims at justifying what was decided, it does not add to the understand of what *wasn't* decided. We aim to offer insights into this area and, specifically, how to nurture the prerequisite conditions for doing so.

In the organisational sensemaking field, the focus of many researchers is mainly on the social process of individual identity in successive spheres of membership through interactions with others. Collective sensemaking (the process by which people give meaning to their collective experiences) does consider the collective (that is, the team) but only with respect to its relationship to the organisation, not to its collective understanding of an external domain. Some researchers, notably Daniel Russell (2009) from a Human-Computer Interaction (HCI) perspective, look at sensemaking for a broader purpose - to collect and organise information in order to gain insight, to analyse, to transfer. However, although his view establishes sensemaking in a collective location (an information world), he describes a style of engagement of sensemaking that is essentially personal, not collective. Of interest in this area is the Cynefin framework (Kurtz and Snowden, 2003) which is a sensemaking framework that is designed to allow shared understandings to emerge which could be insightful with respect to how teams ingest, socialise, and collectively store insights. As with other collective sensemaking models, however, it has resonance in early problem-solving stages and for formal and finite periods of time whereas our interest is on the full product life cycle. Other researchers (Klein et al., 2006; Naumer et al., 2008; Kolko, 2010) elaborate further by bringing data-framing into the picture and defining design synthesis as a process of sense-making, trying to make sense of chaos. The data-framing activity of sensemaking lends itself to being part of a long-term collective effort to understand and therefore may have some relevance to future research building upon our study.

## 6. RESEARCH DESIGN: CONSTRUCTIVIST GROUNDED THEORY

As our primary interest is on substantive theory generation rather than extending or verifying extant theory, we take an interpretive epistemological stance, by adopting the Constructivist Grounded Theory qualitative research methodology described by Charmaz (2014). Grounded Theory is highly applicable in this type of research because the method is explicitly emergent, taking an inductive approach where no adequate prior theory exists. This method is well-suited for a "What is going on here?" type of qualitative inquiry as this study is and where the insights generated will relate to a specific research situation. The Constructivist version of Grounded Theory is especially useful in this study because the researchers bring significant industry experience to bear and the method embraces and manages researcher knowledge and experience as a valuable asset rather than a liability to be minimised.

The Constructivist Grounded Theory method is applicable for this study because both the current Agile paradigm for software development and our area of interest in the discipline of requirements engineering (RE) focuses on people and interactions and, as a qualitative research method, Grounded Theory allows for the study of complex, multi-faceted social interactions and behaviour. The use of Grounded Theory in software engineering and computer science research has risen significantly since 2005 and specifically used successfully to study Agile software development teams, e.g. Adolph et al., 2011; Dagenais et al., 2010; Sedano et al, 2019; Hoda and Noble, 2017; Stol et al., 2016; Stray et al, 2016.

### 6.1 Data collection

We use theoretical sampling (Charmaz, 2014) where the analysis of the data collected prior informs both the selection of, and inquiry with, the next participants, allowing the sample of participants and questions to purposefully evolve as patterns emerge in the data until a theory is reached. Individual participants and corporate sites selected were ones involved with software product development (teams developing software for market as opposed to bespoke solutions either for external or internal clients) and that claimed to have cross-functional product development teams. We purposefully excluded teams involved in developing development tools, dev-ops products, etc., as well as entertainment or personal productivity products. The reason for this is that it was felt that development team members might have a natural understanding of needs in those product areas simply by virtue of their professional expertise or social experience. We selected corporate participants that were software companies developing products for domains quite different than a software development team would naturally be expected to deeply understand beforehand. As our goal was to gain a deep understanding and familiarity with groups and their collective and individual behaviours, our primary data collection methods were observations of team meetings and team interactions, documented by thick descriptions and reflections soon after the meetings. We enriched this data with semi-structured interviews (recorded and transcribed) using open-ended questions to allow real issues to freely emerge. Thus, the method was grounded in the participants' world, with the data interpreted and the emerging and evolving theory constructed by the researcher and the participants.

We engaged with seven software firms, carefully recruited through our professional networks and via direct outreach to select organisations. Five of these firms produce commercial enterprise-class software products, one creates sophisticated virtualisation solutions, while another develops large-scale aerospace systems as bespoke system development. Three of these firms have adopted agile as a paradigm (i.e., guided by agile principles), three approach agile as a structured methodology (e.g., *"we do SCRUM by the book"*), while the other employs a highly prescriptive methodology due to the bespoke nature of its software development and the dictates of its customers. All firms are major players in their markets. The firms range in age from 8 to 50 years and in size from thirteen to several thousand employees (see Figure 1.).

| | markets served | # employees | age of firm (years) | dev. process model |
|---|---|---|---|---|
| 1 | social media marketing | 2,000 | 12 | agile as a methodology |
| 2 | payment solutions | 100 | 25 | agile as a methodology |
| 3 | vertical market virtualisation solutions | 200 | 17 | agile as a methodology |
| 4 | cell therapy lab & clinic mgmt | 25 | 22 | agile as a paradigm |
| 5 | retail | 500 | 21 | agile as a paradigm |
| 6 | enterprise skills management | 13 | 8 | agile as a paradigm |
| 7 | satellite & ground imagery | 3,000 | 50 | prescriptive methodology |

Figure 1. Participant companies

The study had 18 product development teams and 27 individuals in a variety of roles formally participate across these companies.  The teams were both formally and informally observed in action (typically during various forms of planning or design sessions) while the individuals participated primarily via semi-structured interviews.

| # teams | 18 |
|---|---|
| # formal observation sessions | 20 |
| # observation hours (formal + informal) | 72 |
| # individual interviews | 27 |
| senior managers | 2 |
| senior engineers / team leads | 8 |
| intermediate s/w engineers | 11 |
| quality assurance specialist | 1 |
| product managers | 5 |

Figure 2.  Participants teams and individuals

In nearly all cases, we hold interviews in the participant's workplace to allow for record review to enrich the interview data.  Also, having approval from all the organisations involved, we are able to locate ourselves as unobtrusively as possible in the workplace to also allow for direct, informal observation which served as an additional data source to the formal observation sessions as a guide to direct further data collection and analysis.  The interviews conducted with individuals were recorded by prior agreement (which was always granted) and later transcribed.  Group interviews were always an option as an additional data gathering technique, but the need has not arisen.  Group observation sessions are documented immediately after in order to capture the important details and flavour of the session.

*6.2 Data analysis*

Given the interpretive nature of this research method, extensive memoing was conducted throughout, used as an essential analytical and reflexive tool.  Coding of interview transcripts and observation session write ups also fed the memoing.  First cycle coding techniques were primarily descriptive and holistic coding to highlight what was interesting and set the 'noise' aside.  Second cycle coding utilised focussed and theoretical techniques.  The Constant Comparative Method (Charmaz, 2014) was used throughout the study along with much use of the "pawing" technique (Ryan and Bernard, 2003).

We used the NVivo (2018) software tool to aid in the management of the unstructured qualitative data.  Data collection stopped once the analysis indicated the achievement of theoretical saturation, the point at which gathering more data reveals no new properties nor yields any further theoretical insights about the emerging grounded theory (Charmaz, 2014).  Dey (1999) prefers the term "theoretical sufficiency" (p. 257), a term which better fits how grounded theory is conducted since the notion of theoretical saturation is not a precise moment nor an exhaustive finish. What sometimes initially appears as a new insight after further analysis can be seen as a variation on existing data, hence the declaration of theoretical saturation is a judgement call, a judgement about the sufficiency for the research question at hand.  The use of the Grounded Theory Method's 'theoretical saturation' technique ensures a certain degree of consistency in the analysis.

We present more details on the results of the analysis in the next chapter but, as an example of the many analytic paths taken, what follows is a high-level view of how one category emerged:

> Interview quote: (in response to the interviewer's question to a developer about his interactions with UX designers on the team) "…really only to ask detail questions that might come up when I'm writing the code. But I'm careful not to criticise their designs. **They're the experts, I trust them**.".  Seeing echoes of this sentiment in other interviews, we created a *concept* of *showing functional deference within the team*.  Later, as we saw indications of a distinct *lack* of functional deference in some teams, we renamed the concept ***functional deference within the team*** with an intent to examine why it existed strongly in some teams and was notably absent in others.
>
> We also observed development teams interacting with dissimilar functions outside of their team and began to see functional deference in a different dimension, e.g., Researcher's observation note from a product road-map presentation by product management to the development team: "… a rip-roaring debate between the dev team and the PMs about the vision of the product and how it might evolve.  Stark contrast with the passive, near silent, reception by the dev team at GGG in a meeting with a very similar purpose.  This is functional deference (or not, in this case) *but between teams of different roles rather than between roles within a dev team*."
>
> We then created a *category* of **functional deference** which had both intra-team and inter-team sub-categories.  Comparable analytic journeys occurred for other categories which were themed and ultimately led to the phenomenon of **Blurring Boundaries**.

Select codes, categories, and concepts from the grounded theory analysis are highlighted in the remainder of this article as bold italics and bold in quotes where it is an in vivo code or quotation.

Validity and limitations are discussed in Chapter 8 – Discussion.

7.  OBSERVATIONS AND ANALYSIS

Our RQ called for "what factors support or impede cross-functional software product teams in collectively achieving a deep and shared understanding of the product domain?"  In this chapter, we examine the relevant factors arising from our data and analysis and do this through the lens of two contexts within the firm: the organisational model and the product planning process. In the following chapter we discuss themes that emerge.

The first context is the organisational model within which the team resides. We presented earlier results of this in a previous paper (Fuller, 2019a) and we expand upon those results in the next subchapter and further place into context in Chapter 8 - Discussion, Chapter 9 - Implications for Practice, and Chapter 10 - Conclusions and Future Work.

The other context we highlight is the product planning model - the who, what, when, where, and why of the product plans. Specifically, how much CFPTs participate in (and, possibly, are responsible for) the core elements of product visioning, strategic planning, and roadmapping.

These contexts are not mutually independent in that the model for product planning, and its influence on the CFPT, is influenced and shaped by the organisational model but also with its own independent dynamics. Separately and collectively, these two models within a company have a significant influence on the CFPT's motivation to grok the product domain as the foundation for the teams' ability to deeply understand the product requirements and the context of the requirements.

*A. The Organisational Model Impact*

To facilitate the discussion of the Organisation Model and its impact, we will describe two figurative models of teams that we use to describe the influences in this model.

In Figure 3, we describe one end of this team model spectrum where functional departments exist in the organisation that those departments contribute functional specialists to a CFPT. Within the teams, there is a strong recognition of and respect for the functional departments outside the team, illustrated by the distinct colour (functional) boundaries within the team. Surrounding the team, there is an ambiguous boundary, reflecting the differing mandates of the participating functional departments (e.g., server-side engineering, product design, mobile development, product management, architecture, etc.). This extreme form of a team model is a workgroup, which we have labelled an *Assembly of Experts* team.

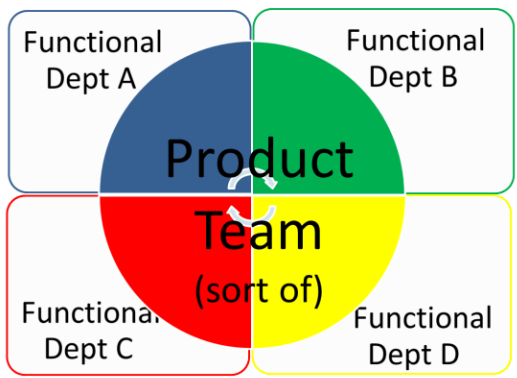

Figure 3. *Assembly of Experts* Team

In Figure 4, we illustrate a contrasting model at the opposite end of the team model spectrum. In this model there are no strong functional home departments for the specialists on the team. The functional distinctions within the team soften as the team works together, determined less by functional competencies. There is a very distinct border surrounding the team, reflecting an unambiguous team mandate. We have labelled this the *True* Team model.

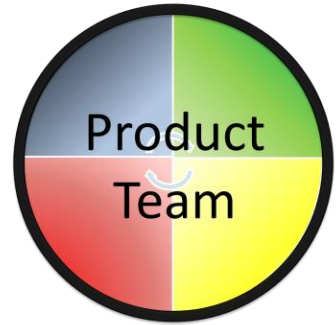

Figure 4. *True* Team

Examining teams in our study against these models, four impacts of note were made contrasting these models:
1. *functional deference* between different functions in the team and the resulting overall **team cohesion**
2. *primary affiliation* of the individual team member
3. *horizon of interest* held by individuals on the team, for the team as a whole, and the **team ownership** of the product overall
4. *alignment with expectations* of the team's degree of product ownership held by senior leadership
5. *individual agenda*

Each of these impacts will be discussed further throughout this chapter.

Note that by deference we are referring to a team or an individual showing submission and respect (yielding) to another team or individual solely due to the other's specific functional expertise being different than one's own.

*1) Intra-team Deference and Overall Team Cohesion*

Do members on a team contribute as being 'on the team' or as being 'an expert in attendance'? We will start this discussion by first looking at *Assembly of Experts* teams (Figure 3).

When a functional organisational structure exists in the software product enterprise, e.g., separate departments (sometimes even sub-departments) for software engineering, product design, product management, each contributing individuals to form CFPTs (see Figure 3), we found that team members are more likely to limit their contributions to the team to topics directly relating to their particular competency and to show marked deference to team members from different functions (**functional deference**) on topics outside their area of primary functional expertise. Although **team cohesion** was influenced by other factors as well, high levels of functional deference appeared to weaken the sense of cohesion in the team. In other words, it was more difficult for an *Assembly of Experts* team to feel united towards a common purpose.

This behaviour showed very strongly in our research. We observed 14 teams in companies with a distinct functional organisation surrounding them and all 14 displayed very strong deference between functions within their teams and exhibiting behaviour as *Assembly of Experts* teams.

Despite a certain ambiguity in this respect, it was clear that the individual sense of *primary affiliation* was stronger toward their functional department than it was toward the software product team and its mission. Simply put, an individual in this organisational model locates themselves via function more than via team membership.

This was unsurprising since, for the teams that more closely resembled the *Assembly of Experts* team model, there is a competing social identity, i.e. the department vs the team. We observed that, in circumstances where there could be conflict between the two, individuals would often feel safer favouring their obligation to their home department. To do otherwise would risk creating inter-departmental tension outside of the team and our observations indicated that departmental connections were often stronger than team connections under these conditions (see Chapter 8 – Discussion regarding *psychological safety* and its role in innovation and team grokking).

In this *Assembly of Experts* model, the enterprise has typically not defined these CFPTs in a manner where an individual can resolve broader organisational conflicts between the different product functions represented on the team. Many enterprises in our study either did not perceive the need to do so or they otherwise concluded that the departmental distinctions were there for good reasons that overrode this consideration. We discuss this point further in Chapter 9 – Implications for Practice.

Illustrative comments of *functional deference* and low *team cohesion* in this team model were frequently heard in the verbal language, For example, when referring to team members in other functional roles, we often heard statements such as, "**I just do my job and they do theirs**", "**They're the experts, I trust them**", or "**I think someone else is looking after that**". This was very strong 'us and them' language, illustrating a strong sense of distinction and separation between the functions within the team (distinct colour boundaries in Figure 3), inhibiting a single sense of team and reflecting the divisions represented by the functional departmental organisation outside the team. This was also illustrated in the body language where it was observed that interactions amongst team members in the same functional roles on teams was often notably more open and exploratory than the more formal interactions between team members in differing functional roles. Similarly, "**I just do what I'm asked to do**" when referring to involvement with the requirements specifications, reflecting a feeling of *playing a role* on the team in contrast with *being* on the team.

While the deference dynamics in this model might be viewed as institutionalised passivity, we saw no evidence that this deference reflected any diminished concern for the quality of the work performed. Rather it was more an acute awareness of one's *primary affiliation* as well as the *horizon of interest* that individuals had. We discuss this latter point further in the next subchapter.

In contrast, turning now to organisations without a functional structure (or at least a weak one) surrounding the CFPTs (see Figure 4), we observed product teams with more *team cohesion* with richer intra-team interactions and softer (sometimes even an absence of) functional interfaces amongst individual team members. This behaviour presented itself strongly - of the 4 teams in companies without a functional organisation, 3 displayed almost no evidence of *functional deference* while the 4$^{th}$ displayed so only mildly, appearing to reflect a very polite and passive corporate culture.

These teams also had much less interaction with, and made less reference to, organisational units outside the team, illustrating a stronger sense of self-sufficiency and *clarity of ownership*. This resulted in markedly less *functional deference* shown on the team and *team cohesion* being notably higher, with individual team members placing the interests of the product foremost and above any functional differences or tensions within the team

When we did observe what might seem as deference being displayed between functions in these teams, it was usually after the team member had fully expressed his/her opinion, typically in the form of a brief qualifying statement at the end, such as e.g. "**however, I'm not a designer**" (or developer, etc.), acknowledging expertise differences and showing respect but not letting the role differences prevent the team member from fully expressing what they felt was best for the product (*team ownership*). Collaborative behaviours were higher in these *True* Team models (Figure 4) and the sense that they were all part of the same team (*team cohesion*) was much stronger. Thus, the language within these teams was much more 'we' oriented and less likely to include connotations of 'us' and 'them' when referring to intra-team functions. One of our participant companies has multiple products in a well-defined product domain and yet it was common that the product team members' LinkedIn profiles showed the product name as the company they work for with little or no reference to the real corporation. They were very clear where they belonged and what they were committed to. (And they appear to smile more).

*2) Playing the Short or the Long Game – Individual and Team horizon of interest and product ownership*

How far into the future do members on the team see themselves as team members? Again, we will start this discussion by first looking at *Assembly of Experts* teams (Figure 3).

Individuals on teams closer to the *Assembly of Experts* model appear to be less invested in the overall success of the product and of the team itself. All 14 teams in companies with a distinct functional organisation surrounding the CFPTs acted in ways that demonstrated strong focus on the near-term roadmap, yet with very little attention given to the long-term plans of the product.

We identified two causes for a limited *horizon of interest* for individuals and for teams. The first, which we discuss here, is the influence the organisational model has on CFPTs in this regard. The second is the influence of the Product Planning Process model and we discuss this in part B of this chapter.

One significant cause for a limited *horizon of interest* is that, when there was a departmental organisational model outside the team, individual team members tended to expect to be reassigned on a more frequent basis. This increased expectation of mobility appeared to reduce an individual's intrinsic connection to the product team and, therefore, the product and the longer-term product roadmap. If a team member expects to be reassigned in the foreseeable future, he/she is likely to hold a certain tentativeness to their commitment to both the product and the product team and are therefore less likely to behave as if they are 'all-in'. Comments heard from team members in this state were typified by: "**I'm on this team… for now**". Sometimes this condition was created or exacerbated when a Human Resources practice was in place that encourages internal mobility within the company (usually in an attempt to foster alignment, engagement, individual growth opportunities, etc.). A telling quote from an engineering manager in one of these situations, "**I don't know how a true 'team' can emerge this way.**", referring to the shorter perspective (*horizon of interest*) that individual team members showed in these conditions and to the increased on-boarding effort required by teams as a result.

When the game was on, team members in the *Assembly of Experts* model tended to take a narrower specialist viewpoint on product issues, reflecting their home department focus. This showed as individual team members (depending on their functional expertise) being much more concerned about how, what OR why a product was to be built but rarely all three, showing *functional deference* to others regarding other perspectives, "**I just do what I'm asked to do**" or "**I do my job and they do theirs**".

Thus, these *Assembly of Experts* teams were generally much less inclined to play the long game. Instead, the focus tended towards the current and near-term plans, "**we just do what the roadmap (or Product Management) says**", since both *functional deference* and a tentative *horizon of interest* undermined a complete commitment to the full product and to the product's long-term development roadmap.

This appeared to affect detailed development decisions as well as we observed, during development discussions and decision-making, less consideration paid to elements of the long-term product roadmap.

With lower *team cohesion* and shorter *horizon of interest*, *Assembly of Experts* teams then were less likely to take collective responsibility for the product's current success or failure. With a (near-term) *narrow lens* on the product plan, they would take ownership for the work they performed, but not the overall product result. They are more likely to passively criticise than attempt to understand the vision and broader plan - "**It isn't the strategy I would have put together**" or "**I don't think they understand the market very well**". The strong 'they' language indicated lack of *ownership* and/or strongly held opinions coupled with a sense of having little control or influence.

In contrast, all 4 teams without a functional organisation surrounding them (Figure 4) regularly positioned their actions in the context of the longer-term plan and vision of the product. These teams tended to exhibit a more complete sense of *team ownership* for their product and a commitment to its longer-term plan (longer *team horizon of interest*), engaging in more comprehensive discussions regarding the product requirements (*striving to grok*). Using a *broader lens*, they strove to understand the *what* of the requirements specifications to a deeper degree and also attempted to understand the *why* of the requirements as best they could. This deeper examination of the context of the requirements is behaviour we would expect to reduce the *requirements fixation* that Mohanani et al. (2014) identified that can occur as a result of requirements framing and our observations support this. Taking this broader perspective, attempting to understand a richer context of the requirements, is behaviour that we would expect to move these teams in a favourable direction with respect to the widespread cause of project failure identified by NaPiRE (Wagner et al., 2019) as being *incomplete requirements*.

In short, regardless of functional role, members of teams in the *True* Team model tended to develop a more holistic and collective perspective of what they were doing (*striving to grok*) and all team members were more likely to care about what, why, and how of the product, reflecting a stronger sense of (collective) *team ownership* ("**it's our product**"), ownership beyond simply their individual contribution to it.

*3) Alignment of Team Ownership of the Product and Senior Leadership Expectation*

In addition to teams closer to the *Assembly of Experts* model not focussing further into the future of the product roadmap than what is deemed minimally necessary to complete current and near-term development, we also observed that these teams are less concerned with whether they are as self-sufficient as they could be, treating structural topics as "**management's problem**". In other words, a view that someone or some group had conscious choices to structure things as they were and the team isn't in a position to change those decisions. Similarly, these teams tend to be less concerned about having limited scope due to organisational or expertise limitations, putting only modest effort into attempting to obtain the necessary skills or resources in order to assume a broader scope or plan. This is an example of the aforementioned point about teams in this organisational model having less ***team cohesion*** and greater ***functional deference*** being less likely to take collective ***team ownership*** for the product and work output. In other words, they are more willing to defer to their respective functional departmental decisioning regarding resourcing, plans, process, etc. than they are to strongly advocate, collectively, for what is in the best interest of the overall product.

While this behaviour was constant across our 14 Assembly of Experts participant teams, we found differences in the understanding held by senior leadership. For 8 of the 14 teams we observed, senior management was generally aware of this (low) level of ***team ownership*** but unaware of its cause. For the other 6 teams in this category, there was a significant disconnect between the low sense of ***team ownership*** of the product and roadmap versus what senior leadership thought was the case.

In all examples of the *Assembly of Experts* teams, senior leadership felt they had assembled these CFPTs with representatives from various functional departments as an expedient way to integrate expertise and to operate across the functions in the organisation for the purpose of addressing the growing complexity and the need for innovation in a highly competitive product world. However, these teams typically behaved as task teams with a short ***team horizon of interest***, focussed almost exclusively on the near-term work in front of them ("**I just do what I'm asked to do**"). In further discussions with these senior leaders about our observations, they sometimes reacted with surprise, stating that they were unaware that these organisational choices would have the impact on individuals and teams that they were having. In a few cases, however, the response was a knowing nod, adding that the condition and solution was beyond their control. We will discuss this point further in Chapter 9 – Implications for Practice.

In contrast, we found that by placing the interest of their product foremost (***team ownership***), *True* Teams were more inclined to advocate for what they felt was in the best interest of the product (***striving to grok***), attaching the team definition of success with the success of the product. As was stated, "**it's our product**". These cases all aligned well with how senior leadership expected the teams to behave.

In an apparent effort to create meaning by collective sensemaking, we observed that most teams, of either model, have a natural propensity to ***want to own*** something and will, therefore, define themselves around what they *can* own. We will come back to this point several times in the remainder of this article.

In summary, these three impacts (***team cohesion***, ***horizon of interest***, and ***alignment with expectations***) illustrate that a CFPT's progression along a spectrum between an *Assembly of Experts* group focussed on the work in front of them and a true empowered, cohesive team committed to the long-term future of the product is heavily influenced by the broader functional organisational structure surrounding the team and, if a functional structure exists, how strong those departmental distinctions are.

*B. The Product Planning Model Impact*

The process for product planning also has a significant impact on cross-functional product teams (CFPTs). Our primary interest was in how (if at all) CFPTs participated in the core elements of product visioning, strategic planning, and roadmapping and how much they appeared to own what they did participate in.

CFPTs that have stronger internal connections and softer functional role deference tend to ask broader questions, be more open-minded, show more curiosity, and attempt to explore more (***striving to grok***). These are all behaviours shown by Mitchell et al. (2009) to be essential ingredients for creativity and innovation. Thus, it would be expected that we observed those teams having more interest in the broader product planning process.

However, our observations also included teams in some environments that did not have the organisational structure and/or culture that allowed teams to own as much of (or even participate in) the product planning process as much as the teams wished they could. This was most often the case where strategic planning for products occurred in a very separate functional area, with the plans communicated, in various manners, to the product development group to be developed. As noted above, we observed that CFPTs ***want to own*** something and would, therefore, *broaden (or narrow) their lens* on the product development plan to match what they were permitted to own.

We use here a concept of ***Broadening (and narrowing) the Lens*** as an explanatory metaphor to describe the mechanism that CFPTs use to adjust the scope and clarity of what they can see of the product domain and what they then subsequently focus on. ***Broadening the Lens*** illuminates a broader picture, with less detail, exposing control boundaries, relationships, and patterns that cannot be seen when the lens is more focussed.

Narrowing the lens tightens the scope, reduces the context, and brings detail into focus.

This action of ***Broadening the Lens*** allows the team to see a bigger picture and, with that broader understanding, the team is then able to more purposefully and knowledgeably re-focus (*narrowing the lens*) to do the work. These are key actions in the team's collective capability to explore further and innovate more. This mechanism also allows teams to make decisions more knowledgeably rather than making safe assumptions and creating, what might have been avoidable, accidental complexity in the process.

The more a team can adjust its lens, the more aligned their definition of success can be with what the company expects from the team (*alignment with expectations*). In the team's attempt to define and honour ownable boundaries, both individuals and teams *colour within the lines* they are given, allowed, or are able to create (*want to own*). The team sense of what their boundaries are is reflected in what completed work the development teams takes pride in and celebrates, e.g. a successful iteration, meeting a release deadline, getting an important feature on the release train, or simply being part of a team that created a product that is successful in the market.

In this context, the potential spectrum that a CFPT is able to navigate within ranges from being spoon-fed tasks for development on one extreme (**"I just do what the story says"**) to having full *team ownership* of visioning, strategic planning, and execution of the product on the other (**"it's our product"**).

As with the Organisational Model, we found that the degree to which a team was able to **Broaden the Lens** (and then narrow it) on their product planning activities also appeared in the verbal language used by the teams. The broader the team's planning scope was, the more we observed conversations indicating a deep understanding of (or, at least references to) product needs with a product domain perspective, product/market opportunities, competitive factors, etc. (*broadening the lens*) and then, with that broader perspective as context, discussing more granular detail as necessary (*narrowing the lens*). Discussions held by teams low on this spectrum referenced domain considerations much less with the conversation almost entirely about internal entities and artefacts such as stories, requirement specifications, other functions/teams, processes, etc.

## 8. DISCUSSION

### A. Positioning

In this chapter, we discuss 3 main themes that emerged in our analysis: 1) the metaphor of blurring boundaries for CFPTs to develop a richer, collective context for understanding product requirements; 2) the contraindication of the 'as machine' metaphor commonly used in organisational and process modelling within software companies; and, 3) the importance of human factors in requirements engineering.

#### 1) Blurring Boundaries

We noted that cross-functional product teams (CFPTs) have a propensity to *want to understand* as much as they are allowed, therefore, it was not surprising to observe teams that had little to no *functional deference* within the team, and who were not being spoon-fed their development tasks, collectively *striving to grok* the product domain as much as they were able and, hence, create a richer collective context of the product requirements.

We also observed that empowered and cohesive CFPTs play a longer *horizon of interest* (they play the long game). With less *functional deference* and less tentativeness of their *horizon of interest* with respect to the individuals' membership on the team, conditions exist that encourage full participation and commitment (both individual and collective) to the long-term product roadmap (***"it's our product"***).

These two observations - Broadening the Lens towards understanding the product domain and Broadening the Lens to own the product plan - are important because all software is developed in context and it is context that guides decisions made throughout the development life cycle. Decisions made by every team member, explicit decisions and tacit ones. If the team is cohesive, their context will be more collective than if it is not (Organisational Model). The more the team owns (or is at least heavily involved with) the product visioning & planning, its context will be more comprehensive than if it does not (Product Planning Process).

Both of these factors, the degree of collectiveness and comprehensiveness of the team's contextual understanding, contribute to the effectiveness of the team's efforts to collectively grok the product domain, its ability and motivation to do so, allowing for a more accurate and comprehensive understanding of the context of the product requirements.

The spectrum of collective domain understanding ranges from not asking questions ("**just do what the story says**") through intellectual domain understanding (learned knowledge of vocabulary, workflows, objectives, etc.) and on to true felt (as if lived) understanding of the domain. The further a team moves along this spectrum the more the team groks. This ***striving to grok*** is a *blurring of the boundaries* between the team and the product domain.

In this context of requirements engineering, we suggest that empathy, specifically collective cognitive empathy, is a fundamentally important ability for the team to possess in order to more deeply understand a domain for which the team is otherwise unfamiliar. Exercising that ability to cognitively empathise, to figuratively step into that other domain, involves a certain temporary softening of the distinction between the team as a collective and the product domain itself. This is what we are calling ***Blurring Boundaries***, temporarily softening the distinctions in order to more truly understand perspectives in another domain. ***Broadening the Lens*** is a necessary technique for the team to be able to see the other domain and its context, and ***Blurring the Boundaries*** is an effort to get closer, to deeply understand (that is, *to grok*). The intergroup theory from Smith et al. (2007) suggests that empathy, as an ability inherently directed at a different individual/group, can become a collective ability and that it can be an attribute of the group that is more than just the aggregation of individuals' attributes. Our observations support that view. We observed that teams that were **Blurring Boundaries** (*striving to grok* the product domain) were the ones closer to the *true* team model and, the closer they were to that model (i.e., the **team cohesion** they had, with clear and broad **ownership**), the more they seemed to be attempting to blur the boundaries in a *collective* manner.

We also observed certain teams appearing to make no attempt to grok the product domain at all, a reflection of the culture of the team and of the organisation. These were all very strong *Assembly of Experts* teams. Certain other teams that did try to grok the product domain had modest success due to influences from their organisational and/or product planning contexts as we discussed in Chapter 7.

To summarise, for a CFPT to be able to figuratively step into another domain and to do so collectively, it is necessary for it to see itself as a cohesive unit (***team cohesion***). This can only be achieved when there is a high level of transparency across all functions on the team, with little to no ***functional deference*** shown within the team, and a strong sense of ***team ownership*** for the product. In other words, a *true* team with a strong product mandate – ***Blurring Boundaries*** with a strong sense of the collective. It requires team members to feel psychologically safe (Edmondson and Lei, 2014; Google re:Work, 2020), have open minds, a high level of curiosity, and a strong common purpose (Mitchell et al., 2009). If any of these are weak or missing, efforts toward discovery, innovation, and collective grokking are all inhibited (Reiter-Palmon & Harms, 2017).

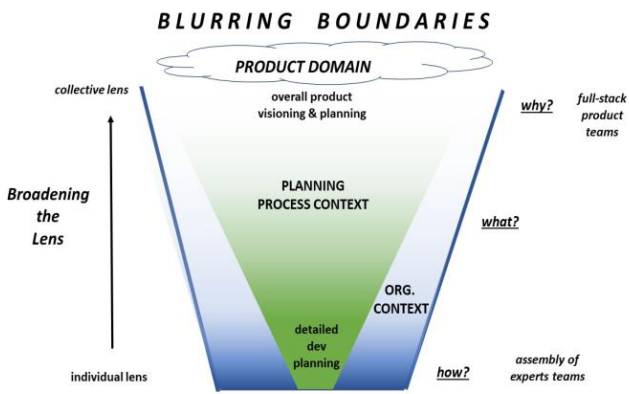

Figure 5 - Blurring Boundaries

*2) "As Machine" models are counterproductive*

Awa (2016) noted that, "[the] functional organisational structure works well in a stable environment where business strategies are less inclined to need changes or updating" (p. 1). However, none of these conditions characterise the software products industry which faces continual change, change due to fashion, technology, and economics, and where competitive forces are often based on innovation and speed to market more than they are on lower cost.

At the team level, studies by Gladstein (2006) and Ancona & Caldwell (2008) indicate that contextual factors (e.g., organisational structure, resources available, and functional mix) have a greater influence on team effectiveness than do internal team processes. We add that the two are not unrelated – our results show that the operating context of a team has a significant impact on internal team character and behaviour including, but not limited to, processes.

Reiter-Palmon et al. (2012) stated, "results indicate that functional diversity is positively related to creativity and innovation" (p. 298). This would appear to support efforts to create cross-functional product teams (CFPTs) even in companies with an overall functional organisational structure.

With this thought, and exploring Awa's point further, we asked the senior leaders about their intent when forming their CFPTs, whether they were formed as special cases or whether creating them indicated that the main development activities were no longer well-contained within a functional area. They generally responded that they had simply formed CFPTs with an instinctive belief that it would improve both innovation and productivity and also that it would simplify project management. On the surface, this looks like a tactic to counteract the negative consequences of the functional organisational structure. However, this is very often not what they are achieving.

Although, as we noted earlier, studies have shown that team goals which emphasise collaboration, open discussion and inclusive team behaviours are critical for team creativity and innovation (e.g., Mitchell et al., 2009), Awa (2016) goes on to point out that, "[the] challenge of the functional structure is the tendency for employees to take a specialist viewpoint" (p. 2) (***functional deference***). We observed that strong ***functional deference*** inhibits high-bandwidth communication across functional groups within the team (a core principle of the Agile paradigm) and segments goal setting, thus interfering with team creativity and innovation. It also makes the team less collectively intelligent. Woolley et al (2010) found that collective intelligence was positively correlated with equal distribution of conversational turn-taking.

In short, our results support the view that functional diversity has a positive effect if the functions are cohesive (i.e., working together as a *True* Team) as opposed to simply working side-by-side (as in the *Assembly of Experts* style teams). Put another way, functional diversity in action adds to creativity and innovation, not simply functional diversity in appearance. Companies with functional organisational structures are often not getting the results they hoped for when they formed their CFPTs.

Productivity is also negatively impacted when teams are viewed as inorganic 'Lego blocks'. Tuckman (1965) described a Forming-Storming-Norming-Performing model of team development evolution, the implications of which are that too much change to team composition while in the norming or performing stages of team development places backward pressure towards the storming or even the forming stages. With some teams in our study, several factors frustrated their efforts to become a team (frequent team member reassignment, corporate human resources rotation policy, high turnover levels, etc.), resulting in the teams spending too little time in the Forming or Storming stages compared to time spent in the Performing stage. Those finding themselves perpetually in the Forming or Storming states remain an Assembly of Experts workgroup, possibly with functional sub-groups inheriting similar traits.

As organisations scale, whether at the enterprise or department level, there is an unconscious tendency towards even more specialisation and any divisions and boundaries that already exist tend to become even more pronounced and hardened. In the software industry this is often intentional since the industry (and the technology space, in general) is dominated by analytical minds, minds that build things, and the things they build are like machines, machines with specialised components that have precisely defined roles. Thus, we often see organisational and process models defined in software organisations using an *as machine* industrial metaphor, and it appears that sometimes these are created without question. This is not unlike the way that early heavily prescriptive SDLCs reflected industrial manufacturing metaphors that eventually came to be generally considered ill-suited to a rapidly changing and increasingly complex software product world. We argued (2019b) that this Organisation as Machine industrial metaphor has outlived its appropriateness, particularly as it applies to requirements engineering (RE), and we now make similar arguments regarding the product planning process when implemented in a mechanistic, as a machine, type model. Software products RE, with its multi-layered tacit dimensions and *other world* target, cannot be adequately addressed by a mechanistic metaphor.

*3) RE is a Team Sport and Human Factors are Key*

NaPiRE notes that after 40 years of research and practice, the discipline of RE still struggles, despite significant contributions made over those years in the form of tools, methods, and process (Wagner et al. 2019). As noted earlier, NaPiRE surveys indicate that many organisations in industry adopt these RE methods and even they still report the most common cause of project failure being incomplete requirements. Our results lead us to the conviction that this will continue to be true until more attention is given to the fundamental issues of how to get tacit knowledge into the software product development teams, i.e. how to help those teams grok their remote and complex product domains. This cannot be achieved simply by improving requirements elicitation and validation approaches that rely on better analytical techniques, and more comprehensive templates while depending on the completeness and accuracy of the 'whisper-game' like communications along a development life cycle process chain. Until the human factors in RE, which NaPiRE acknowledges is a major characteristic of the RE discipline (Wagner et al., 2019), receives more attention, RE will continue to struggle in the respect of inadequate requirements understanding.

*B. Quality and Validity*

Quality in Constructivist Grounded Theory research is assessed in terms of validity and transferability which, together, determine some measure of usefulness, supported by evidence of credibility, originality, and resonance (Charmaz, 2014).

During the research, we employed various strategies (Maxwell, 2012) to mitigate threats to validity, as well as taking actions to allow an assessment of the degree of transferability of the results.

First, we have intensive, ongoing involvement with the individual and organisational participants. This extended participation also includes the capability to 'live' in the participants' workplace, observing and interacting, formally and informally.

It was through this intensive involvement that all data was collected and reflections made. This ensured that the data is grounded in the experiences of a large number and range of participants and settings, providing us with richer types of data, data that is more direct (less dependent on inference). This involvement also permits repeated observations and interviews as well as a convenient opportunity to re-examine and review observations and analysis with participants, which helps rule out spurious associations, systematic biases, and premature theories.

Triangulation, data collected from a range of participants and settings, further reduces the risk of chance associations and systematic biases. The diversity in participants includes a range of corporate types (size, age, markets), teams, and types of individuals (experience levels and functional roles).

We collected and used rich data (transcribed interviews and thick, descriptive notetaking of observations) that provided a fuller and more revealing picture of what is going on. All interviewees are offered, as part of the standard closing of the interview, an opportunity to review the transcription once it was complete. None have chosen to, which we take as trust in the interviewer and process. However, many expressed a strong interest in seeing the final research results when they are made available, expressing a keen interest in the topic and the potential usefulness of the results - an early hint of resonance.

Due to the interpretive nature of Constructivist Grounded Theory and the role the researcher perspective plays, replication of results is inherently difficult. However, we are ensuring auditability by maintaining detailed records of data collected from which one can see the general stability of the coding over time.

Participant (and non-participant) checks are conducted periodically throughout the study to obtain reactions on both the emerging analysis and conclusions drawn. This offers us multiple perspectives on the analysis and it also helps to rule out possibilities of misinterpretation. No feedback has yet been received that is contrary to what the results indicate.

We are assessing the transferability of the results within the context of software product development primarily via peer reviews (reviews with software product development leaders) of the resulting theory. Later, we intend to draw comparisons with non-product software development teams to further refine the specificity of transferability claims.

We have very long and deep professional experience in the software product industry and we recognised at the outset of this research that this strong positionality shapes our objectivity and subjectivity of many aspects of perspective in this study. While we acknowledge the challenges, we consider this experience, and the bias it creates, to be an asset to this research. As pointed out in Chapter 6, Constructivist Grounded Theory embraces researcher experience as a manageable asset to the research. As Eisner (1998) suggested, the expert ability to "see what counts" – the sensitivity to tacit elements of the data, meanings, and connotations – guides our research, supported fully by the collected data, towards questions and insights that matter. Simply put, the research team groks the world of the participants. Also, our many years of experience with the same types of people that are participants affords us considerable comfort, understanding, and rapid rapport with the participants which results in dialogue that would have been very difficult to achieve by researchers lacking this level of industrial experience.

## 9. Implications for Practice

As noted in Chapter 3, a core motivation for this research was to create insights that would be helpful to industry practitioners managing the entire life cycle of developing and managing software products. In addition to this motivation giving rise to the RQ, it also guided our choice of method, sampling of participants, and validation of results. We present the following implications of our results for industrial practitioners, primarily senior leaders in software product companies, in the form of factors to be aware of and recommendations to consider with the hope that increasing the awareness of the impacts of certain software product development leadership practices will help organisations and teams move towards achieving a deeper and more collective understanding of product requirements. While some of what follows may read as conventional wisdom, we found that conventional wisdom and conventional practice are often disconnected.

### Implication 1 – Organisational Design

There is a significant difference between a product team member feeling a primary sense of affiliation to a product, being part of the product team, bringing their particular competencies to that team versus an individual feeling primarily affiliated with their functional group or department within the company and being assigned to bring their skill to a particular team. What T-shirt would your back-end software engineer wear - a functional group T-shirt with a specific product badge on it, or specific product team T-shirt, possibly with badge indicating their primary competency on the team (Fuller, 2009a)? While the words may sound subtle the impact is significant.

Recommendation: eliminate (at least minimise) any ambiguity a team member may have regarding primary affiliation. This is important in order to create cohesive teams that have minimal intra-team deference across functions, higher bandwidth communications, and more unified goal setting, all leading to greater creativity and innovation. There are other ways to address knowledge management, functional leadership, and career development consideration without encouraging individual affiliation around their functional expertise.

### Implication 2 – Team Membership

Changes in team membership are sometimes necessary for a variety of reasons (skills adjustment, inter-personal considerations, etc.), however unnecessary team membership changes come with a price tag. Having individuals and, by extension, the entire team being committed to the long-term vision and plan for the product is in part dependent upon an expectation that each team member is on the team for the long term.

Recommendation: make team composition changes only when necessary and when the benefit to the team justifies the price the team will pay for the change.

### Implication 3 – Team Mandate

Work groups are content to consume a short-term work plan. Conversely, cohesive and healthy teams want to own something for the longer term and these teams will define ownable boundaries for themselves if those boundaries are not already specified.

Recommendation: provide the product development teams unambiguous ownership of a product or subset of a product. Define that ownership in as meaningful terms as possible, something that the team can feel inspired by and feeling proud to own.

### Implication 4 – Collective Grokking of the Product Domain

Cohesive teams also want to deeply understand what they own. If they are not being spoon-fed tasks or having the product domain interpreted for them, they will try to collectively grok it themselves.

Recommendation: expect the product development teams to deeply understand the product domain and ensure that the teams are aware of this expectation.

### Implication 5 – Team involvement in Product Visioning & Planning

Much is lost when the product development teams do not participate in the higher-level product visioning and planning, determining the *whats* and *whys* behind the requirements. Teams that are less involved in these activities, make more assumptions (many of them tacit), create more accidental complexity, and take more time. In short, run a greater risk that the work product will fail to meet the desired outcome. Additionally, this limited involvement results in an erosion of deep ownership. Product development teams that do not own (or meaningfully participate) in product visioning and planning are less likely to take collective responsibility for the product success or failure.

Recommendation: give the product development teams as much ownership as possible for product visioning and planning. At the very least have them deeply involved in all key product visioning and planning activities.

## 9. Conclusions and Future Work

### Conclusions

Our RQ asked - *"what factors support or impede cross-functional software product teams in collectively achieving a deep and shared understanding of the product domain?"*. This question is important because the success of software products and the software companies themselves is impacted by how well software development teams deeply understand the context of the product requirements. Software development leaders have no theories to guide them in this regard.

Our results point to two factors that address the RQ. Both the organisational model surrounding the team and the team's role in the product planning model have significant impacts on team dynamics, product ownership, time horizon, and level of collective grokking of the product domain. We identified ***blurring boundaries*** as a basic metaphor to describe what teams are doing in their effort to grok, i.e. to deeply understand the context of product requirements by imaginatively stepping into that other domain. This purposeful blurring occurs, to varying degrees, within the teams, between the team and the product planning process, and between the team and the product domain.

Aspects of our results sit in opposition to existing software development organisational and process models and views commonly found in many technology companies which encourage specialisation and clearly defined segregation of duties. This specialisation is intended to support knowledge management, ensure specialised functional expertise in senior leadership, support career advancement for highly specialised roles, provide focus to meet deadlines, and to fit with adopted process models. Many senior leaders in our study had an instinctive belief that productivity and innovation would be maximised through specialisation, yet evidence indicates quite the opposite. It shows that, for software product organisations, both productivity and innovation are impeded by these hardened, sharpened, and specialised boundaries and methods.

We conclude that the multi-disciplinary creativity and innovation necessary to create software products in a complex and uncertain world calls for a rethinking of the software product development organisation, a softening of distinctions, a ***blurring of boundaries***, the antithesis of a mechanistic, *as machine*, metaphoric view.

### Future Work

This research shows that a cross-functional product team's ability to *blur the boundaries* both within the team, within the company, and between the team and its product domain generally defines the team's capacity to collectively grok. We saw that, at the very least, this ability is influenced by the organisational and product planning process models in the company. However, we acknowledge that there may well be additional factors at play that call for further exploration. For example, writing this at the midst of the COVID-19 crisis and being acutely aware of the topic of working location, it is clear that fully co-located vs partially distributed vs fully distributed product development teams is a topic that calls for examination, specifically as it relates to this research question.

There is also further exploration needed to better understand the factors at play regarding a cross-functional software product team's definition of team success. We observed empowered teams being committed to the long-term success of the product while teams closer to an assembly of experts model tended to shrink the boundaries of ownership until they had reached a space for which they could claim ownership. It is unclear if this simply reflects a difference in ownership scope or if these two team models define success using more fundamentally different considerations.

This research raises questions regarding education. For example, how are technically-oriented people (primarily millennials) working in teams (typically cross-functional) and following a rational process to create software solutions able to develop, nurture, and incorporate 'squishier' skills into a process that strives to be as rational and deterministic as possible? Are current educational curricula adequate?

We found evidence that striving to be rational and deterministic is an unquestioned given in some firms. With statements from some senior executives in our study that the organisational model and inter-departmental mandates were beyond their individual influence, it appears that, in many organisations, the *as a machine* metaphor often may be more deeply rooted. De Alencar (2012) found "despite an awareness of the need for creativity and innovation for organizational success, deep rooted tendencies to maintain the status quo prevailed, making it difficult to introduce changes in a direction of promising conditions for creativity". With creativity and innovation being imperatives for the success of software product companies, this calls for further exploration.

Finally, while we observed teams using the ***broadening the lens*** mechanism to ***blur boundaries*** between the team and the product domain, there remains an important and challenging question to be answered and that is why, *even in the same organisational, process, and leadership environment,* some software product teams observably achieve deeper grokking of the product domain than do others. We believe this remains an important area of future inquiry for both academia and industry.

ACKNOWLEDGMENT

This research was funded in part by the Natural Sciences and Engineering Research Council of Canada (NSERC).